\documentclass{PoS}

\usepackage{psfig}

\PoS{PoS(BDMH2004)005}

\title{Disk galaxy evolution up to redshift z=1
\thanks{Based on observations with the European 
Southern Observatory Very Large Telescope (ESO-VLT), 
observing run IDs 65.O-0049, 66.A-0547 and 68.A-0013;
and observations with the NASA/ESA \emph{Hubble Space Telescope},
PID 9502.}}

\ShortTitle{Disk galaxy evolution up to redshift z=1}

\author{\speaker{Asmus B\"ohm} and Bodo~L.~Ziegler\\ 
        Universitätssternwarte Göttingen, Germany\\
        E-mail: \email{boehm@uni-sw.gwdg.de}, 
	        \email{bziegler@uni-sw.gwdg.de}}

\abstract{
We have performed intermediate-resolution VLT/FORS spectroscopy and
HST/ACS imaging of 129 field spiral galaxies within the FORS Deep Field. The
galaxies cover the redshift range $0.1 \le z \le 1.0$ and comprise all types 
from Sa
to Sdm/Im. Spatially resolved rotation curves were extracted and fitted
with synthetic velocity fields that take into account all geometric (e.g.,
inclination and misalignment) and observational effects (in particular,
blurring due to optical beam smearing and seeing). Using these fits, the
maximum rotation velocity $\vm$ could be determined for 73 objects.

The Tully-Fisher relation of this sample at a mean look-back time of $\sim$\,5
Gyr shows a luminosity evolution which amounts to $\sim$\,2\,mag in 
rest-frame $B$ for low-mass spirals ($\vm$\,$\approx$\,100\,km/s) but is 
negligible for high-mass spirals ($\vm$\,$\approx$\,300\,km/s). 
This confirms our previous analysis which was limited to ground-based imaging.
The observed overluminosity of low-mass galaxies is at variance with
predictions from simulations. On the other hand, at given $\vm$, we find slightly 
smaller disk sizes towards higher redshifts, in compliance with the CDM 
hierarchical model. The observed mass-dependent luminosity evolution might 
therefore point towards the need for a more realistic modelling of the stellar 
(i.e. baryonic) component in $N$-body codes.
}

\FullConference{Baryons in Dark Matter Halos\\
                 5-9 October 2004\\
                 Novigrad, Croatia}

\newcommand{\vm}{V_{\rm max}}

\newcommand{\rd}{r_{\rm d}}
\newcommand{\dmb}{\langle M_B \rangle}

\begin{document}

\section{Introduction}

The Cold Dark Matter hierarchical scenario has become one of the paradigms in
astrophysics and cosmology. On scales of clusters of galaxies
and beyond, the observed structures are very well
reproduced by simulations that assume a flat $\Lambda$CDM cosmology. 
However, the observed properties of
\emph{individual} galaxies remain challenging to the models.
For example, semi-analytic recipes fail to reproduce the blue colors of
low--mass spirals and the red colors of high--mass spirals in the local
universe (Bell et al. 2003). To gain further insight into this issue,
we performed an observational study of distant field galaxies
with a data set that probes more than half the age of the universe.
Utilising the Tully--Fisher Relation (TFR) between luminosity and
maximum rotation velocity $\vm$ 
as well as the velocity--size relation
between $\vm$ and disk scale length $\rd$ we will quantify the
evolution of late--type galaxies over the past 8\,Gyr.

Throughout this article,
the concordance cosmology with
$\Omega_{\rm m}$ = 0.3, $\Omega_\Lambda$ = 0.7 and 
$H_0$ = 70\,km\,s$^{-1}$\,Mpc$^{-1}$ has been assumed.

\section{The Sample}

Our data set has been selected within the FORS Deep
Field (FDF, see Heidt et al.~2003), a multi--band photometric survey
performed with the Very Large Telescope and the
New Technology Telescope operated by ESO.
Based on a catalogue with spectrophotometric types and photometric redshift 
estimates, we chose objects for follow--up spectroscopy
basically upon a late--type Spectral Energy Distribution 
and apparent $R$-band magnitude $R\le23$\,mag.
In total, we took spectra of 129 galaxies with the 
FORS1 \& 2 instruments of the VLT.
For an accurate derivation of the galaxies' disk inclinations, scale lengths
etc.~we also obtained Hubble Space Telescope images of the FDF using the 
Advanced Camera for Surveys (F814W filter).

We extracted spatially
resolved rotation curves by fitting Gaussians to the usable emission lines
stepwise along the spatial axes
of the spectra. Due to the small apparent sizes of the galaxies, the slits
used for spectroscopy covered substantial fractions of the disks. 
This ``optical beam smearing'' and the seeing resulted in a heavy blurring
of the observed rotation curves. To overcome these effects, we generated
synthetic velocity fields that introduced the same observational effects
as the data and used these to model the observed rotation curves. With
this strategy, the \emph{intrinsic} maximum rotation velocity $\vm$ was
derived for 73 galaxies within the field--of--view of the ACS images.
Out of these, 34 rotation curves robustly probed the region of
constant rotation velocity and had a high degree of symmetry; 
these will be referred to as high quality data in the following. 
The kinematic sample with 73 objects spans a redshift range $0.09 \le z \le 0.97$ 
corresponding to look-back times
$1.2\,{\rm Gyr} \le t_1 \le 7.6\,{\rm Gyr}$ with a median
$\langle t_1 \rangle = 4.7$\,Gyr. 

\section{Discussion \& Conclusions}

\begin{figure}[t]
\centerline{
\hspace{-1cm}
\psfig{file=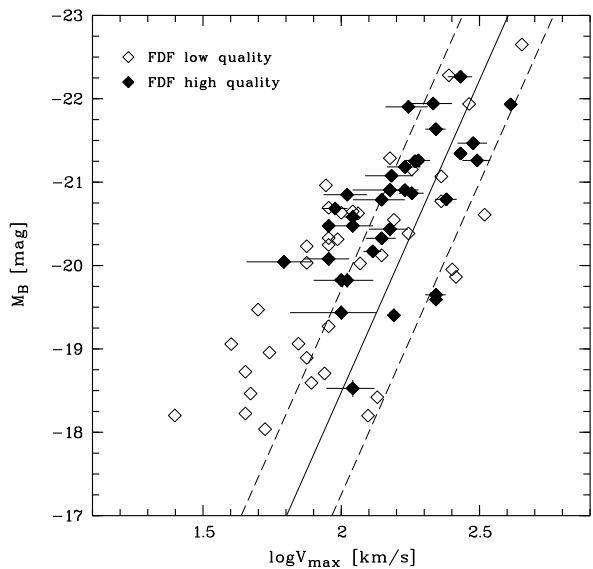,width=7.8cm,angle=-90}
}
\caption{\label{fdfpt}
FORS Deep Field sample of spirals in the range $0.1 \le z \le 1.0$ in 
comparison to the local Tully--Fisher relation  
by Pierce \& Tully (1992, solid line; 
dashed lines give 3\,$\sigma$ limits). The distant sample is
subdivided into high quality rotation curves (solid symbols) and low quality
rotation curves (open symbols).
Error bars are shown for the high quality data only.
}
\end{figure}

\begin{figure}[t]
\centerline{
\hspace{-1cm}
\psfig{file=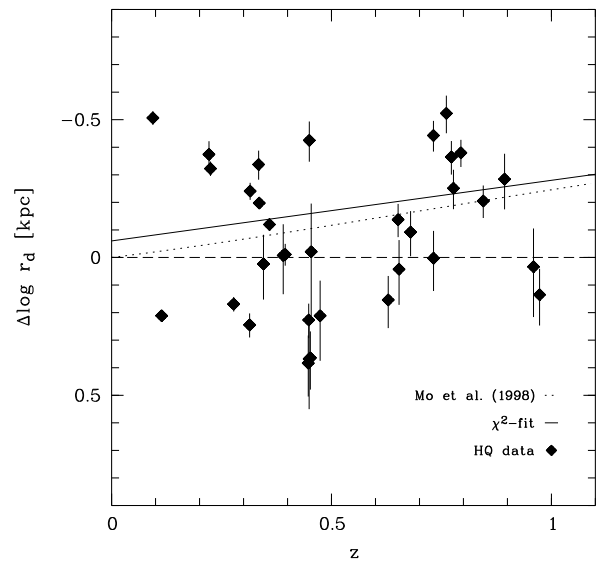,width=7.8cm,angle=-90}
}
\caption{\label{fdfvsr}
Offsets $\Delta \log \rd$
of the distant FORS Deep Field galaxies with high quality rotation curves
from the local $\vm$--$\rd$ relation (reference: Haynes et al.~1999) 
as a function of redshift. 
Objects with $\Delta \log r_{\rm d}>0$ have \emph{larger} disks than local spirals at 
a given $\vm$, whereas values $\Delta \log r_{\rm d}<0$ correspond to disks 
which are \emph{smaller} than in the local universe. As indicated by the fit
to the data (solid line), we find a slight trend towards smaller disks at
higher redshifts, in agreement with theoretical predictions 
based on the Cold Dark Matter hierarchical scenario 
(Mo et al.~1998, dotted line).
}
\end{figure}

In Fig.~\ref{fdfpt}, we compare the FDF spirals at 
$\langle z \rangle \approx 0.5$ to the TFR of local spirals as given by 
Pierce \& Tully (1992). At fixed $\vm$~--- which corresponds to a fixed
total mass~---, the distant galaxies are more luminous than their $z \approx 0$
counterparts. For the total sample and the high quality data,
the average offsets are $\dmb = -0.98$\,mag and $\dmb = -0.81$\,mag,
respectively. These overluminosities may indicate 
decreased $M/L$ ratios due to, e.g., younger stellar populations in
the distant galaxies.
Furthermore, we find evidence for a differential evolution:
while the \emph{massive} distant spirals are in relatively well agreement with
the local TFR, the \emph{low--mass} distant galaxies are increasingly overluminous
towards low $\vm$. Using a parameterisation $M_B = a \log \vm + b$ for the TFR,
a bootstrap bisector fit with 100 iterations yields
a slope $a=-4.05 \pm 0.58$ for the high quality data, significantly shallower
than the local slope $a=-7.48$. Since the analysis presented here
is based on the ACS imaging, this confirms the results shown in
B\"ohm et al.~(2004) which were limited to ground--based imaging.
Our data may thus indicate a mass--dependent luminosity evolution that would
be at variance with simulations: e.g., Steinmetz \& Navarro
(1999) found a constant slope with their SPH code, while Boissier \& Prantzos
(2001) even predict a steepening of the TFR towards longer look-back times.
We performed numerous tests to rule out the possibility of an incompleteness
bias or systematic error in our analysis 
(see B\"ohm et al.~2004). Since the intermediate--redshift disks
are smaller than in the local universe (Fig.~\ref{fdfvsr})~--- 
as is to be expected in a
cosmology with hierarchical structure growth~--- the
evolution of spiral galaxies we observe would be at variance with theoretical
predictions only in terms of the \emph{stellar population properties}.

An analysis of the broad--band colors of our sample galaxies with single--zone
models of chemical enrichment showed that the low--mass FDF spirals began
to turn their gas into stars at later cosmic epochs and on longer timescales
than the high--mass spirals (see Ferreras et al.~2004). When evolved to
zero redshift, the model stellar populations of low--mass galaxies have
younger mean stellar ages and a broader age distribution than those of
high--mass galaxies.
Although these were relatively simple models without any spatial resolution,
the result may point towards an \emph{anti--hierarchical evolution} of
the baryonic component in late--type galaxies which is also known as
``down--sizing''.

\end{document}